\documentclass[12pt]{iopart}

\usepackage{iopams}  	
\usepackage{float}     
\usepackage{graphicx}
\usepackage{xcolor}
\definecolor{darkblue}{rgb}{0.0, 0.0, 0.55}
\usepackage{adjustbox}
\usepackage{hyperref}
\hypersetup{colorlinks=true, linkcolor=darkblue,filecolor=darkblue, urlcolor=darkblue,}
\begin{document}
\title{Entanglement protection in higher-dimensional systems}
\author{Ashutosh Singh$^*$, and Urbasi Sinha$^{**}$}
\address{Light and Matter Physics Group, Raman Research Institute, Sadashivanagar, Bangalore-560080, India.}
\ead{$^*$asinghrri@gmail.com, $^{**}$usinha@rri.res.in}
\vspace{10pt}
\begin{indented}
\item[]\date{\today}
\end{indented}

\begin{abstract}
The inevitable dissipative interaction of an entangled quantum system with its environment causes degradation in quantum correlations present in the system. This can lead to a finite-time disappearance of entanglement, which is known as Entanglement Sudden Death (ESD). Here, we consider an initially entangled qubit-qutrit system and a dissipative noise which leads to ESD, and propose a set of local unitary operations, which when applied on the qubit, qutrit, or both subsystems during the decoherence process, cause ESD to be hastened, delayed, or avoided altogether, depending on its time of application. Delay and avoidance of ESD may find practical application in quantum information processing protocols that would otherwise suffer due to short lifetime of entanglement. The physical implementation of these local unitaries is discussed in the context of an atomic system. The simulation results of such ESD manipulations are presented for two different classes of initially entangled qubit-qutrit systems. A prescription for generalization of this scheme to a qutrit-qutrit system is given. This technique for entanglement protection in the noisy environment is compared with other related techniques such as weak measurement reversal, dynamic decoupling, and quantum Zeno effect.
\end{abstract}

\section{Introduction}

Quantum decoherence [\ref{1}] is a ubiquitous and unavoidable phenomenon arising because of  entanglement between quantum systems  and their environment. Entanglement, on the other hand, is a fundamentally important phenomenon in the studies of quantum foundations and has great significance in quantum technologies. It is  now seen  as  an  indispensable resource  in Quantum  Information Processing  (QIP) for  various tasks such  as   quantum  computation,  teleportation,   superdense  coding, cryptography, etc., which are either impossible or less efficient using classical correlations [\ref{2}-\ref{4}].

The inevitable dissipative interaction  of an entangled quantum system
with its environment leads to  an irreversible loss of single particle
coherence  as well  as degradation  of entanglement  present in  the
system [\ref{5}]. For  some initial  states, in  the presence  of an
Amplitude Damping Channel (ADC), entanglement degrades asymptotically,
whereas for  others it  can disappear  in finite  time, also  known as
early stage disentanglement or Entanglement Sudden Death (ESD), in
literature [\ref{6}-\ref{8}]. Soon  after its theoretical prediction,
ESD was experimentally demonstrated  in atomic [\ref{9}] and photonic
[\ref{10}]  systems. The  real world  success of several quantum
information,  communication,  and  computation tasks  depends  on  the
resilience of entanglement from noises present in the environment, and
longevity of the entanglement. Thus,  ESD poses a practical limitation
on QIP  tasks.  Therefore,  strategies  which   make entanglement robust against  the detrimental effects of  the noise are
of practical interest in QIP.

Owing to the  simplicity of two-qubit entangled systems and its  usefulness as a resource  in QIP,  there have  been several  theoretical proposals  to combat decoherence  and finite-time  disentanglement in  these systems [\ref{11}-\ref{24}].  Some of these proposals have been experimentally demonstrated   in   atomic,  photonic,  and   solid   state   systems [\ref{25}-\ref{34}]. One of the entanglement protection schemes [\ref{21}] considered a class of two-qubit entangled states which undergo ESD in the presence of an ADC. For such systems, a Local Unitary Operation (LUO), in this case the Pauli $\sigma_x$ operator also known as NOT operation, has been proposed. When LUO is applied  on one or both the subsystems during  the process of decoherence, it can  hasten, delay, or completely avoid the ESD, all depending on the time when NOT operation is applied. This   proposal  was  later  transformed by our group  into  an  all-optical experimental  setup  to study  the  effect  of  LUOs on the disentanglement dynamics in  a photonic system [\ref{22}]. This work proposed  an experimentally  feasible architecture for the implementation of ADC and the local  NOT operation to suitably manipulate the ESD of a two-qubit system in a controlled manner.

Higher-dimensional entangled quantum systems (qudits, $d\geq 3$) can offer practical advantages over the canonical two-qubit entangled systems in QIP protocols. These  systems are more  resilient to errors than their qubit counterparts  in quantum cryptography, and they offer practical advantages; for example, increased channel capacity in quantum communication,  enhanced security in QIP  protocols, efficient
quantum  gates,  and  in  the  tests of  the  foundations  of  quantum
mechanics [\ref{35}]. It  is  therefore  important  to  study  the  effects  of decoherence on these systems.  Entanglement evolution has been studied in higher-dimensional systems present  in the noisy  environment, and despite the  resilience of these  systems to noise [\ref{36}],  ESD is established to be  ubiquitous in all dimensions of  the Hilbert spaces [\ref{37}, \ref{38}]. 

As a first step towards generalization to higher dimensional systems, a qubit-qutrit system is the natural choice as it is intermediate in complexity to a qubit-qubit and qudit-qudit system. Given the availability of full-fledged separability criteria for pure as well as mixed state qubit-qutrit systems, experimental feasibility of state preparation, and the significance of higher-dimensional systems in QIP, they become an important architecture towards understanding and controlling the effects of decoherence in higher-dimensional systems. Some   attempts  have  been  made   to  locally manipulate  the  disentanglement    dynamics    in    qubit-qutrit [\ref{39}, \ref{40}], and qutrit-qutrit [\ref{41}-\ref{43}] systems to retain the entanglement for longer duration. It is found that the quantum interference between two  upper levels  of  the qutrit  can  also be  used  to control  the disentanglement dynamics in higher-dimensional systems [\ref{39}, \ref{44}].

Consider a practical scenario where Charlie prepares a bipartite entangled state for some QIP task and he has to send the entangled particles to Alice and Bob through a quantum channel which is noisy and can potentially cause disentanglement before the particles reach the two parties. In this scenario, we ask the following question: given a higher-dimensional bipartite entangled state which would undergo ESD in the presence of an ADC, can we alter the time of disentanglement by some suitable LUOs during the process of decoherence? As a possible answer to this question, we  explore   the  generalization  of  the   proposal  in  Ref.~[\ref{21}]  for  higher-dimensional  systems,  say qubit-qutrit or qutrit-qutrit system, in  the presence  of an  ADC. If such systems undergo ESD, we propose a set of LUOs, such that when they are applied on the subsystems during  the  process   of  decoherence, they manipulate the disentanglement dynamics, in particular,  delay the  time at which ESD occurs.  Such a  study was  partially done  in Ref.~[\ref{39}], and  here we  generalize their scheme  and propose  a more general class of  LUOs, which  can always suitably manipulate ESD  for an arbitrary initially  entangled state. In the case of qubits, Ref.~[\ref{21}] found it to be NOT operation which is operation corresponding to the Pauli $\sigma_x$ operator. In this work, for qutrits, we propose a set of LUOs, which allow flipping the population between different  levels of the qutrit. Depending on the combination of LUOs, and the time  of their  application, this  method is shown  to be  able to hasten, delay or completely avoid the ESD in qubit-qutrit as well as qutrit-qutrit systems. 

In  the  decoherence process,  pure  states  evolve to  mixed  states.
Therefore, a  computable measure  of entanglement  for mixed  state is
required to quantify  the entanglement. Negativity is  such a measure,
based on the  Positive Partial Transpose (PPT) criterion  due to Peres
and  Horodecki [\ref{45}, \ref{46}]. For $2\otimes2$  and $2\otimes3$ dimensional systems, all PPT  states are separable and  Negative Partial Transpose (NPT) states are entangled.  Negativity is defined as the sum of the absolute values of all the  negative eigenvalues of the partially transposed density matrix  with respect to one of  the subsystems. While the enigmatic features of entanglement were predicted back in 1935, its witnessing and quantification continues to be at the forefront of current research. For a specific class of higher dimensional states $\rho$ which are supported on $d\otimes D$ Hilbert space ($d \leq D$) and with rank $r(\rho) \leq D$, PPT criterion has been found to be necessary and sufficient condition for separability [\ref{47}]. But for more general higher-dimensional systems ($d \otimes D,~\forall~d, D\geq 3~ \&~r(\rho)>D$), PPT criterion is only  a necessary but not sufficient condition for separability.

For these systems, we study Negativity  Sudden Death  (NSD), whose  non-occurrence guarantees Asymptotic Decay of Entanglement (ADE). But for $3\otimes 3$ dimensional PPT states with $r(\rho) > 3$ which have zero Negativity, we cannot comment on the separability and we need some other measure. For this purpose, we have used matrix realignment method for detecting and quantifying entanglement [\ref{48}-\ref{50}] after state is found to have zero negativity using PPT criterion. Realignment criterion states that for any separable state $\rho$, the trace norm of the realignment matrix $\langle m|\otimes\langle\mu|\rho^R|n\rangle\otimes|\nu\rangle=\langle m|\otimes \langle n|\rho|\mu\rangle\otimes|\nu\rangle$ never exceeds one. Therefore, if the realigned negativity given by $R(\rho)= max(0,||\rho^R||-1)$, is non-zero then state $\rho$ is entangled, where $\rho^R_{ij,kl}=\rho_{ik,jl}$. This criterion can detect some of the bound-entangled states which may not be detected by PPT criterion. We use PPT as well as realignment criterion for entanglement detection.

The rest of the paper is  organized as follows: In section  (\ref{S2}), we briefly discuss the physical model in  which qubit-qutrit entangled system can be realized  in an atomic  system and the  natural presence of  ADC in
such  a system. Next, Kraus  operators governing the evolution  of
quantum system  in the presence of ADC  are given. We then propose a set  of LUOs for qubit, and qutrit for manipulating the ESD, and
their physical  implementation for an  atomic system is  discussed. In
section  (\ref{S3})  and  (\ref{S4}), we  present  our  calculations
implementing  the  proposed  LUOs  on  two different class  of qubit-qutrit  system and  their results.  In section (\ref{S5}), we briefly comment on  the generalization of this proposal
to  higher  dimensions, taking an  example of a two-qutrit system. In section (\ref{S6}), the results of ESD manipulations are
compared and contrasted for a given  initial state with respect to the
choice of various LUOs. In the end, we  conclude with  the  advantages  and limitations  of  our proposal for ESD manipulation with other existing schemes.

\section{Physical Model} \label{S2}
\begin{figure}[!htb]
\begin{center}
\includegraphics[clip, trim=5.25cm 12.4cm 8cm 12.6cm, width=0.75\linewidth]{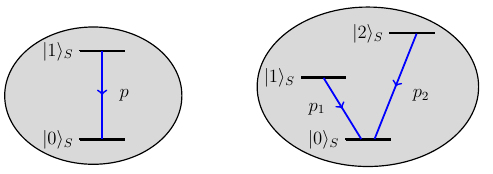}
\caption{\textit{Two-level atom (qubit), and a three-level atom (qutrit) in V-configuration. The parameters $p$, $p_1$, and $p_2$ denote the decay probabilities between the levels $|1\rangle_S\rightarrow|0\rangle_S$ of the qubit, and $|1\rangle_S\rightarrow|0\rangle_S$, and $|2\rangle_S\rightarrow|0\rangle_S$ levels of the qutrit, respectively. The qutrit is taken to be in V-configuration such that the dipole transitions are allowed only between the levels $|1\rangle_S\leftrightarrow|0\rangle_S$ and $|2\rangle_S\leftrightarrow|0\rangle_S$.}}
\label{fig1}
\end{center}
\end{figure}

Consider a hybrid qubit-qutrit entangled system in the presence of an ADC. The qubit and qutrit systems can be realized by a two-level, and a three-level atom in V-configuration (see Fig.~\ref{fig1}), respectively, in an optical trap for instance. The qubit-qutrit entangled state can be prepared by methods discussed in Refs. [\ref{51}-\ref{55}]. Afterwards, the entangled atoms are put into two cavities separated far from each other - by distances larger than the wavelength of the photons emitted from these atoms. The cavities are taken to be at absolute zero temperature and in vacuum state. The spontaneous emission due to atoms interacting locally with their cavities with electromagnetic field in vacuum state, forms identical but independent ADC.

The evolution of a qubit in the presence of an ADC is given by the following quantum map:
\begin{equation}
\eqalign{
|0\rangle_S|0\rangle_E & \rightarrow|0\rangle_S|0\rangle_E~,\cr
|1\rangle_S|0\rangle_E & \rightarrow \sqrt{1-p}|1\rangle_S|0\rangle_E+\sqrt{p}|0\rangle_S|1\rangle_E~,}
\label{eq1}
\end{equation}
where $|0\rangle_S$ and $|1\rangle_S$ are the levels of qubit, and $|0\rangle_E$ and $|1\rangle_E$ are vacuum state and single-photon Fock state of the environment (cavity), respectively. In the Born-Markov approximation: $p=1-\exp(-\Gamma t)$, which is the probability of de-excitation of the qubit from the higher level $|1\rangle_S$ to the lower level $|0\rangle_S$. The subscripts `S' and `E' refer to the system and environment, respectively.

The quantum map of a qutrit (in $V$-configuration ) in the presence of an ADC is given by
\begin{equation}
\eqalign{
|0\rangle_S|0\rangle_E & \rightarrow|0\rangle_S|0\rangle_E~,\cr
|1\rangle_S|0\rangle_E & \rightarrow \sqrt{1-p_1}|1\rangle_S|0\rangle_E+\sqrt{p_1}|0\rangle_S|1\rangle_E~,\cr
|2\rangle_S|0\rangle_E & \rightarrow \sqrt{1-p_2}|2\rangle_S|0\rangle_E+\sqrt{p_2}|0\rangle_S|1\rangle_E~,}
\label{eq2}
\end{equation} 
where $|0\rangle_S$, $|1\rangle_S$, and $|2\rangle_S$ are three levels of the V-type qutrit. There are two decay probabilities for such a qutrit corresponding to the transitions $|1\rangle_S\rightarrow |0\rangle_S$ and $|2\rangle_S\rightarrow |0\rangle_S$, and given by $p_1=1-\exp(-\Gamma_1 t)$ and $p_2=1-\exp(-\Gamma_2 t)$, respectively. Here, $\Gamma_1$ and $\Gamma_2$ represent the decay rate of levels $|1\rangle_S$, and $|2\rangle_S$, respectively.

The Kraus operators governing the evolution of the system in the presence of ADC, for qubit ($M_i$) and qutrit ($\mathbb{M}_i$), are obtained by tracing over the degrees of freedom of the environment from Eqs.~(\ref{eq1}) and (\ref{eq2})~, respectively, and given by
\begin{equation}
\eqalign{
M_0=\left( \begin{array}{cccc}
1 & 0  \cr
0 & \sqrt{1-p}\end{array} \right),~~
M_1=\left( \begin{array}{cccc}
0 & \sqrt{p} \cr
0 & 0 \end{array} \right).}
\label{eq3}
\end{equation}

\begin{equation}
\eqalign{
\mathbb{M}_0=\left( \begin{array}{cccc}
1 & 0 & 0  \cr
0 & \sqrt{1-p_1} & 0\cr
0 & 0 & \sqrt{1-p_2}\end{array} \right),~~\cr
\mathbb{M}_1=\left( \begin{array}{cccc}
0 & \sqrt{p_1} & 0  \cr
0 & 0 & 0 \cr
0 & 0 & 0\end{array} \right),~~
\mathbb{M}_2=\left( \begin{array}{cccc}
0 & 0 & \sqrt{p_2}  \cr
0 & 0 & 0\cr
0 & 0 & 0\end{array} \right).}
\label{eq4}
\end{equation}

The Kraus operators for qubit-qutrit system ($M_{ij}$) are obtained by taking appropriate tensor products of the qubit and qutrit Kraus operators as follows:
\begin{equation}
M_{ij}=M_i \otimes \mathbb{M}_j~~;~~i=0,1,~\&~j=0,1,2.
\label{eq5}
\end{equation}

If initial state of the system is $\rho(0)$ then evolution in the presence of ADC is given by,
\begin{equation}
\rho(p)=\sum_{i,j} M_{ij}~ \rho(0) ~ M_{ij}^\dag.
\label{eq6}
\end{equation}

If such a system undergoes ESD, then manipulation of ESD is achieved by applying LUOs (NOT operations) on qubit and/or qutrit at appropriate time $t=t_n$ as shown in Fig.~\ref{fig2} below. 

\begin{figure}[H]
\begin{center}
\includegraphics[clip, trim=0cm 20cm 1.5cm 1.5cm, width=0.75\linewidth]{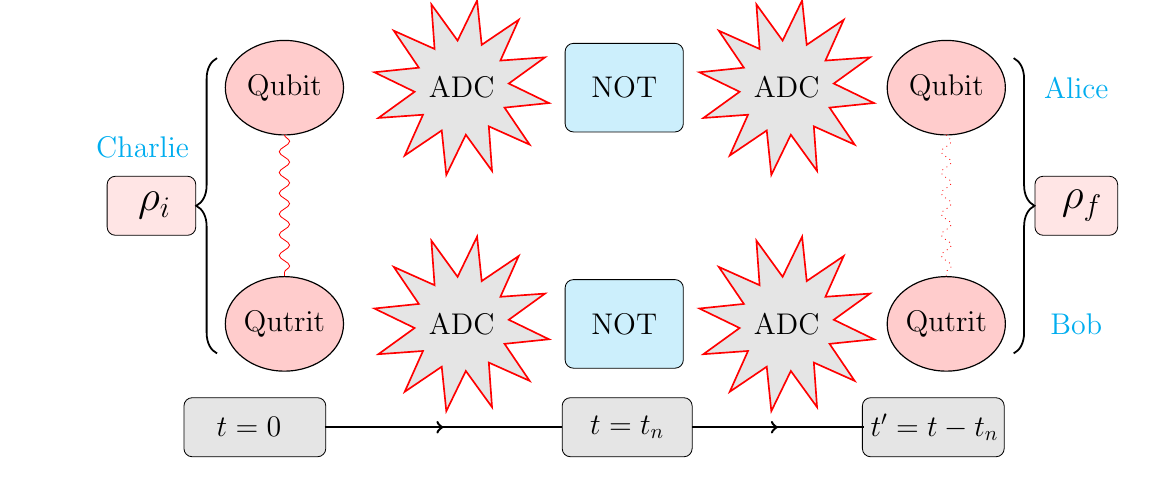}
\caption{\textit{Scheme for prolonging the entanglement in the presence of ADC. Initial state $\rho_i$ undergoes ADC decoherence then NOT operations are performed at $t=t_n$ and system is left to evolve in the ADC. The clock measuring the time is reset at $t=t_n$ and time after the application of NOT operation is measured by $t'=t-t_n$. Such a splitting of time is essential for later discussion.}}
\label{fig2}
\end{center}
\end{figure}

The NOT operation on qubit is Pauli spin operator $\sigma_x$. For qutrit, we propose here a set of operations which swap the populations between the two, or all the three levels of qutrit, namely trit-flip operations: $\mathbb{F}_{01},\mathbb{F}_{02}$, $\mathbb{F}_{102}$, and $\mathbb{F}_{201}$, where
\begin{equation} 
\eqalign{
\mathbb{F}_{01}=\left( \begin{array}{cccc}
0 & 1 & 0 \cr
1 & 0 & 0 \cr
0 & 0 & 1\end{array} \right),~~
\mathbb{F}_{02}=\left( \begin{array}{cccc}
0 & 0 & 1 \cr
0 & 1 & 0 \cr
1 & 0 & 0\end{array} \right),~~\cr
\mathbb{F}_{102}=\left( \begin{array}{cccc}
0 & 1 & 0  \cr
0 & 0 & 1  \cr
1 & 0 & 0\end{array} \right),~~
\mathbb{F}_{201}=\left( \begin{array}{cccc}
0 & 0 & 1  \cr
1 & 0 & 0  \cr
0 & 1 & 0\end{array} \right).}
\label{eq7}
\end{equation}

In an atomic system, the NOT operation on the qubit ($\sigma_x$) can be applied by a $\pi$-pulse on the transition $|0\rangle_S \leftrightarrow |1\rangle_S$, which  interchanges the population between the levels $|0\rangle_S$ and $|1\rangle_S$ of the qubit. The trit-flip operation $\mathbb{F}_{01}$ ( $\mathbb{F}_{02}$) on the qutrit can be applied by a $\pi$-pulse on the transitions $|0\rangle_S \leftrightarrow |1\rangle_S$ ( $|0\rangle_S \leftrightarrow |2\rangle_S$) of the qutrit, which interchanges the population between the respective two levels. The trit-flip operation $\mathbb{F}_{102}$ ($\mathbb{F}_{201}$) on the qutrit can be realized by a $\pi$-pulse applied on the transition $|1\rangle_S \leftrightarrow |0\rangle_S$ ( $|2\rangle_S \leftrightarrow |0\rangle_S$) followed by another $\pi$-pulse to interchange the populations between $|0\rangle_S$ and $|2\rangle_S$ ($|0\rangle_S$ and $|1\rangle_S$). That is, by a series of two $\pi$-pulses $\pi^{|1\rangle_S\leftrightarrow|0\rangle_S}\pi^{|0\rangle_S\leftrightarrow|2\rangle_S}$ ($\pi^{|2\rangle_S\leftrightarrow|0\rangle_S}\pi^{|0\rangle_S\leftrightarrow|1\rangle_S}$).

Mathematically, the application of LUOs at $p=p_n$ can be represented by
\begin{equation}
\rho(p_n)=(U_1\otimes U_2)\rho(p)(U_1\otimes U_2)^\dag~,
\label{eq8}
\end{equation}
where $U_1=\sigma_x$ or $\mathbb{I}_2$, and  $U_2=\mathbb{F}_{01},~\mathbb{F}_{02},~\mathbb{F}_{102},~\mathbb{F}_{201}$ or $\mathbb{I}_3$.

Let us label another set of Kraus operators ($M'_{ij}$) with the parameter $p$ replaced by $p'$ ($t$ replaced by $t'$, $t'=t-t_n$); $p'=1-\exp(-\Gamma t')$, and of the form similar to Eq.~(\ref{eq5}). These Kraus operators are applied to the state (\ref{eq8}) to see the evolution of the system when it undergoes ADC after the application of LUOs as follows:
\begin{equation}
\rho(p',p_n)=\sum_{i,j} M'_{ij}~\rho(p_n)~ M_{ij}^{'\dag}. 
\label{eq9}
\end{equation}

When both the LUOs are identity operations, i.e., $U_1=\mathbb{I}_2$ and $U_2=\mathbb{I}_3$, we have the uninterrupted system evolving in the ADC. The state of the system in this case is given by
\begin{equation}
\rho(p',p)=\sum_{i,j}M'_{ij}\rho(p)M_{ij}^{'\dag}.
\label{eq10}
\end{equation}

Our aim is to investigate whether the phenomenon of hastening, delay and avoidance of ESD also occurs in higher-dimensional entangled systems as seen in $2\otimes 2$ systems [\ref{21}] or not. For this purpose, we compare the Negativity of the system manipulated using LUOs (\ref{eq9}) with that of the uninterrupted system (\ref{eq10}). For the purpose of current analysis, we choose the decay probabilities for the qutrit as $p_1=0.8p$ and $p_2=0.6p$, where $p$ is the decay probability of the qubit.


\section{X-type qubit-qutrit entangled system: State-I}\label{S3}

Let us consider the one-parameter qubit-qutrit entangled state given by,
\begin{equation}
\eqalign{
\rho(0) =& \frac{x}{2}[|00\rangle\langle 00|+|01\rangle\langle 01|+|11\rangle\langle 11|+|12\rangle\langle 12|] \cr
& + \frac{1-2x}{2}[|02\rangle\langle 02|+|02\rangle\langle 10|+|10\rangle\langle 02|+|10\rangle\langle 10|]~,}
\label{eq11}
\end{equation}
where $0\leq x<1/3$.

Assuming that both the subsystems, qubit as well as qutrit, undergo local and independent ADC, evolution of the system is given by Eq.(\ref{eq6}). For state (\ref{eq11}), the mathematical form of Negativity in the presence of ADC and in terms of $x$ and $p$ is given as

\begin{equation}
\begin{adjustbox}{max width=0.8\columnwidth}
$
\eqalign{
N(x,p)= & \frac{1}{4} [2 p^2 x-4px+1.6 p+2 x-(0.64 p^4 x^2-1.28 p^3 x^2+10.24 p^2 x^2+2.56 p^3 x \cr
& -12.16 p^2 x+4.96 p^2-25.6 p x^2+25.6 p x-6.4 p+16 x^2-16 x+4)^{1/2}].}
$
\end{adjustbox}
\label{eq12}
\end{equation}

\begin{figure} [H]
\begin{center}
\includegraphics[width=0.45\linewidth]{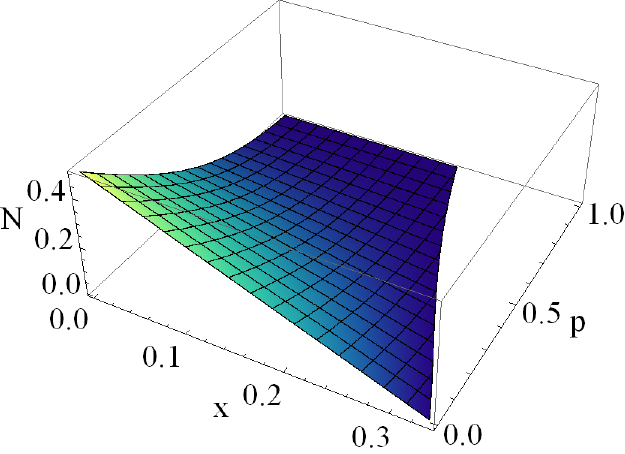}
\caption{\textit{Plot of Negativity vs. ADC probability for  $0\leq x < 1/3$ for the  entangled state (\ref{eq11}). The system undergoes ADE for $0\leq x\leq 0.20$, and ESD for $0.20<x<1/3$.}}
\label{fig3} 
\end{center}
\end{figure}

Evolution of the entanglement vs. ADC parameter ($p$) for $0\leq x<1/3$ is shown in Fig.~\ref{fig3}. The entangled state (\ref{eq11}) undergoes ADE for $0\leq x\leq 0.2$, and ESD for $0.2<x<1/3$. This can be easily verified by Eq. (\ref{eq12}).

\subsection{ESD in the presence of ADC}

We choose $x=0.25$ such that the initial state (\ref{eq11}) undergoes ESD at $p=0.6168$. The mathematical form of Negativity in terms of $p$ and $p'$ is given as

\begin{equation}
\begin{adjustbox}{max width=0.8\columnwidth}
$
\eqalign{
N(p,p')=&\frac{1}{4} [0.34 p^2 p'^2-0.84 p^2 p'-0.84 pp'^2+0.5 p p'+0.5 p'^2+0.6 p'+0.5 p^2+0.6 p+0.5\cr
& -\{0.0256 p^4 p'^4-0.1152 p^4 p'^3+0.1936 p^4 p'^2-0.144 p^4 p'-0.1152 p^3 p'^4+0.0096 p^3 p'^3 \cr 
& + 0.8656 p^3 p'^2-1.32 p^3 p'+0.1936 p^2 p'^4+0.8656 p^2 p'^3-0.9676 p^2 p'^2-2.584 p^2 p' \cr 
& -0.144 p p'^4-1.32 p p'^3-2.584 p p'^2+6.48 p p'+0.04 p'^4+0.56 p'^3+2.56 p'^2-1.6 p' \cr
& +0.04 p^4+0.56 p^3+2.56 p^2-1.6 p+1\}^{1/2}].}
$
\end{adjustbox}
\label{eq13}
\end{equation}

\begin{figure} [H]
\begin{center} 
\includegraphics[width=0.45\linewidth]{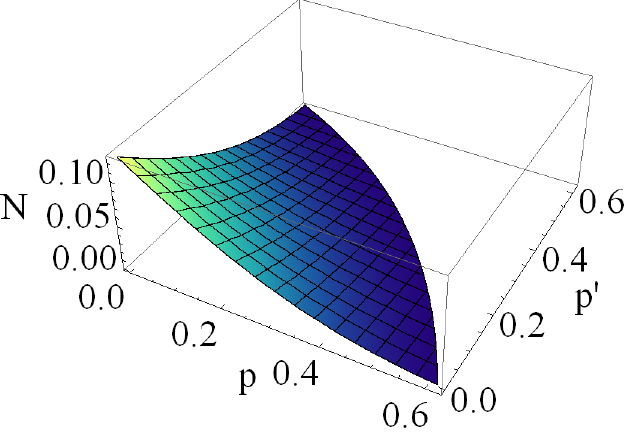}  
\caption{\textit{Plot of Negativity vs. ADC probability ($p,p'$) for $x=0.25$ for the state (\ref{eq11}). It undergoes ESD at $p=0.6168$ for $p'=0$ and vice-versa. For a non-zero $p$, ESD occurs along the curved path in the ($p,p'$) plane.}}
\label{fig4} 
\end{center}
\end{figure}

Using Eq.~(\ref{eq13}), the plot of Negativity vs. ADC probability ($p,p'$) for the state (\ref{eq11}) is shown in Fig.~\ref{fig4}. For $p=0$, ESD occurs at $p'=0.6168$ and for arbitrary values for $p$, ESD occurs along the non-linear curve in $pp'$ plane as shown in Fig.~\ref{fig4}.

For entanglement protection, we apply different combinations of LUOs and their effects are discussed below.

\subsection{$\sigma_x$ applied to qubit and $F_{01}$  applied to qutrit}

A NOT operation ($\sigma_x$) is applied to the qubit and trit-flip operation $\mathbb{F}_{01}$  applied to qutrit part of the state (\ref{eq11}) at $p=p_n$ as in Eq.~(\ref{eq9}).

For uninterrupted system, end of entanglement $p'$ depends on $p$ as follows: 

\begin{equation}
\begin{adjustbox}{max width=0.8\columnwidth}
$
\eqalign{
p'=& \frac{1}{2 (0.0625 p^2-0.15 p+0.0875)}[0.15 p^2+0.035 p-0.25 + 
\{(-0.15 p^2-0.035 p+0.25)^2 \cr
& -4 (0.0625 p^2-0.15 p+0.0875)(0.0875 p^2+0.25 p-0.1875)\}^{1/2}].
}
$
\end{adjustbox}
\label{eq14}
\end{equation}

When LUOs are applied at $p = p_n$ , the end of entanglement $p'$ depends on $p_n$ as follows:
\begin{equation}
p'=\frac{\left(p_n-1\right) \left(0.0875 p_n^2+0.25 p_n-0.1875\right)}{\left(0.25 p_n+0.5\right) \left(0.35 p_n^2+p_n+0.25\right)}.
\label{eq15}
\end{equation}

\begin{figure} [H]
\begin{center}
  \includegraphics[width=0.45\linewidth]{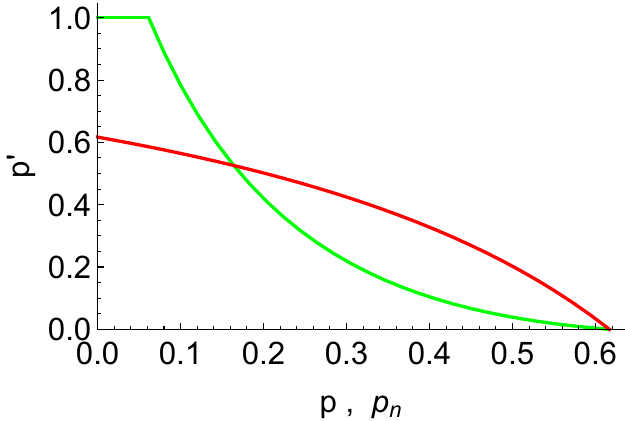}
\end{center}
\caption{\textit{Plot shows the non-linear curvature in  $p'$ vs.  $p ~or~p_n$ for ESD (red curve)  and its manipulation (green curve) such that the NOT operation ($\sigma_x$) is applied on the qubit and trit-flip operation $\mathbb{F}_{01}$ applied on the qutrit at $p=p_n$ for  $x=0.25$. The action of LUOs give rise to avoidance  for $0\leq p_n \leq 0.0615$, delay for $0.0615<p_n<0.1641$, and hastening of ESD for $0.1641<p_n<0.6168$ as the green curve lies below red curve in this range.}}
\label{fig5}
\end{figure}

 The Fig.~\ref{fig5} shows the non-linear curvature in  $p'$ vs.  $p~or~p_n$ for ESD (red curve) and its manipulation (green curve). The manipulation leads to avoidance of ESD for $0\leq p_n \leq 0.0615$, delay for $0.0615<p_n<0.1641$, and hastening of ESD for $0.1641<p_n<0.6168$ as the green curve dips below red curve in this range.  This can be easily verified through Eqs. (\ref{eq14}) and (\ref{eq15}).

\subsection{$F_{01}$ applied to qutrit only}

\begin{figure}[H]
  \centering
  \includegraphics[width=0.45\linewidth]{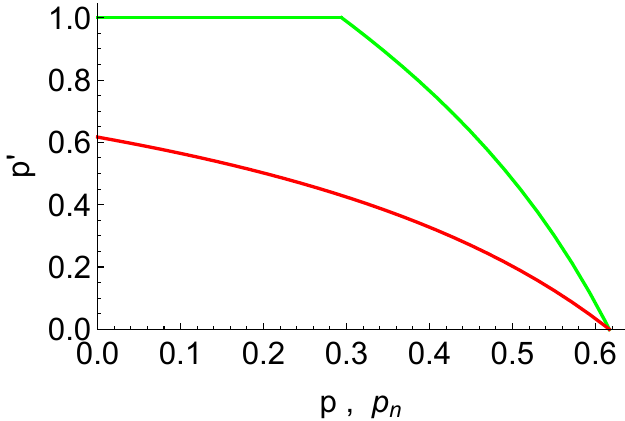}
\caption {\textit{Plot shows the non-linear curvature in  $p'$ vs.  $p ~or~p_n$ in ESD (red curve) and its manipulation (green curve) such that trit-flip operation $\mathbb{F}_{01}$ is applied on the qutrit at $p=p_n$ for $x=0.25$. The action of LUOs give rise to avoidance for $0\leq p_n\leq 0.2941$, delay for $0.2941<p_n<0.6168$, and hastening of ESD  does not occur is this case.}}
\label{fig6}
\end{figure}

The trit-flip operation $\mathbb{F}_{01}$ is applied to the qutrit part of the state (\ref{eq11}) at $p=p_n$ as in Eq.~(\ref{eq9}). The Fig.~\ref{fig6} shows the non-linear curvature in  $p'$ vs.  $p~or~p_n$ in ESD (red curve) and its manipulation (green curve). The manipulation leads to avoidance for $0\leq p_n\leq 0.2941$, delay for $0.2941<p_n<0.6168$, and hastening of ESD  does not occur in this case.

The LUOs $\sigma_x\otimes\mathbb{F}_{102}$ and $\mathbb{I}_2\otimes\mathbb{F}_{102}$ applied on the state (\ref{eq11}) lead to same effect as $\sigma_x\otimes\mathbb{F}_{01}$ and  $\mathbb{I}_2\otimes\mathbb{F}_{01}$, respectively. The other combination of LUOs such as $\sigma_x\otimes\mathbb{F}_{02}$,~$\sigma_x\otimes\mathbb{F}_{201}$,~ $\sigma_x\otimes\mathbb{I}_2$, ~$\mathbb{I}_2\otimes\mathbb{F}_{02}$, and ~$\mathbb{I}_2\otimes\mathbb{F}_{201}$  applied on the state (\ref{eq11}) give rise to only hastening of ESD in the entire range $0< p_n <0.6168$.

Let us now intuitively understand the disentanglement dynamics of the qubit-qutrit system and the occurrence of ESD. The state~(\ref{eq11}) is entangled due to the coherence terms $\rho_{34}$ ($|02\rangle\langle 10|$) and $\rho_{43}$ ($|10\rangle\langle 02|$) of the density matrix. The separability condition depends on the instantaneous value of the quantity $N=\frac{1}{2} \left(\rho_{11}+\rho_{66}-\sqrt{(\rho_{11}- \rho_{66})^2+4 \rho_{34} \rho_{43}}\right)$. If $N$ is negative, the state~(\ref{eq11}) is entangled else separable. In the presence of an ADC, coherence terms $\rho_{34}$ and $\rho_{43}$ decay as $\sim\sqrt{(1-p)(1-p_2)}$ and the term $\rho_{66}$ decays as $\sim(1-p)(1-p_2)$. The population of qubit-qutrit ground state $\rho_{11}$ changes as $\rho_{11} + \rho_{44} p + \rho_{22} p_1 + \rho_{55} p p_1 + \rho_{33} p_2 + \rho_{66} p p_2$. The terms $\rho_{34}$ ($\rho_{43}$) and $\rho_{66}$ decrease with time and $\rho_{11}$ increases. Due to the cumulative evolution of all these terms, the time $\left( t=-\frac{1}{\Gamma} \log_e(1-p)\right)$ at which $N$ becomes zero, is known as the time of sudden death and the state~(\ref{eq11}) becomes separable afterwards. For $p = 1$ or $p_2 = 1$ ($t\to \infty$) system in Eq.~(\ref{eq11}) looses the coherence completely and for $p,~p_1,~p_2 = 1$ qubit-qutrit system is found in the ground state $|00\rangle$.

The physical reason behind the action of different LUOs resulting in hastening, delay, or avoidance of ESD for this class of state can be understood as follows: when we apply the trit-flip operation  $\mathbb{F}_{01}$ (for example) on the system~(\ref{eq11}) after it has evolved in the ADC, it changes the instantaneous population between different levels of the qutrit in such a way that the elements of the density matrix (i) $\rho_{11}$ and $\rho_{22}$, (ii) $\rho_{44}$ and $\rho_{55}$ get swapped, and (iii) the coherence term $\rho_{34}$ ($\rho_{43}$) change their position to $\rho_{35}$ ($\rho_{53}$). When this flipped state evolves in the ADC, new coherence terms $\rho^{(n)}_{35}$ and $\rho^{(n)}_{53}$ (where terms with superscript `(n)' indicate the density matrix elements after the application of LUOs) now decay as $\sim(1-p)(1-p_1)(1-p_2)$, and new $\rho^{(n)}_{66}$ term decays as $\sim (1-p)(1-p_2)$.  The term $\rho^{(n)}_{11}$ evolves as $\rho^{(n)}_{11} + \rho^{(n)}_{44} p + \rho^{(n)}_{22} p_1 + \rho^{(n)}_{55} p p_1 + \rho^{(n)}_{33} p_2 + \rho^{(n)}_{66} p p_2$. After the application of LUOs, condition for ESD now becomes: $N=\frac{1}{2} \left(\rho^{(n)}_{11}+\rho^{(n)}_{66}-\sqrt{\left(\rho^{(n)}_{11}- \rho^{(n)}_{66}\right)^2+4 \rho^{(n)}_{34} \rho^{(n)}_{43}}\right)$, which looks similar to the earlier condition for ESD but it depends on new terms of the density matrix after the LUOs. The instantaneous population of different levels of the qubit-qutrit system depends on the decay rate of the different levels of the qubit-qutrit system, the time when LUO (trit-flip operation) is applied, and time elapsed after the application of LUO. Again, due to cumulative evolution of different terms of the density matrix, when $N$ becomes zero, system becomes separable. The basic idea behind this entanglement protection scheme is to choose correct combination of LUOs depending on the decay rate of different levels of the qubit-qutrit system and the time of application of LUOs such that the state after the application of LUOs results in the delay or avoidance of ESD.


\section{X-type qubit-qutrit entangled system: State-II}\label{S4}

Let us consider another class of one-parameter qubit-qutrit entangled state given by,
\begin{equation}
\eqalign{
\rho(0)= & \frac{x}{2}[|00\rangle\langle 00|+|01\rangle\langle 01|+|11\rangle\langle 11|+|12\rangle\langle 12|+|00\rangle\langle 12| 
+|12\rangle\langle 00|] \cr 
& + \frac{1-2x}{2}[|02\rangle\langle 02|+|10\rangle\langle 10|]~,}
\label{eq16}
\end{equation}
where $1/3<x\leq 1/2$.

Assuming that both the subsystems; qubit as well as qurit, undergo local and independent ADC, evolution of the system is given by Eq.~(\ref{eq6}). For state (\ref{eq16}), evolution of the entanglement vs. ADC parameter ($p$) for $1/3\leq x\leq 1/2$ is computed. The initial entangled state (\ref{eq16}) undergoes ESD in the entire range $1/3<x\leq 1/2$.   For $p=0$, ESD occurs at $p'=0.8452$ and for arbitrary values for $p$, ESD occurs along the non-linear curve in $pp'$ plane similar to Fig.~\ref{fig4}.  We choose $x=0.5$ such that ESD occurs at $p=0.8452$, and study the effect of NOT operation ($\sigma_x$) applied to the qubit, and/or  trit-flip operations $\mathbb{F}_{01}$, $\mathbb{F}_{02}$, $\mathbb{F}_{102}$, $\mathbb{F}_{201}$  applied to the qutrit, in manipulating the ESD. The effect of different combinations of LUOs in entanglement protection are discussed below.

\subsection{$\sigma_x$ applied to qubit and $F_{01}$ applied to qutrit}

\begin{figure} [H]
\centering
\includegraphics[width=0.45\linewidth]{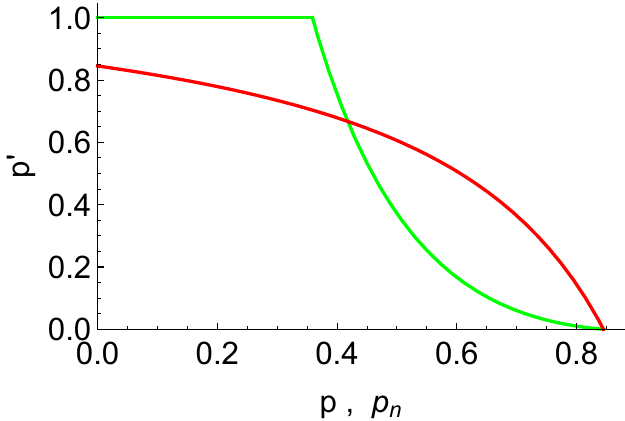}
\caption{\textit{Plot shows the non-linear curvature in  $p'$ vs.  $p ~or~p_n$ for ESD (red curve)  and its manipulation (green curve) such that the NOT operation ($\sigma_x$) is applied on the qubit and trit-flip $\mathbb{F}_{01}$ applied on the qutrit at $p=p_n$ for  $x=0.5$. The action of LUOs give rise to avoidance for $0\leq p_n\leq 0.3586$, delay for $0.3586<p_n<0.4177$, and hastening of ESD for $0.4177<p_n<0.8452$ as green curve lies below the red curve in this range.}}
 \label{fig7}
\end{figure}

The NOT operation ($\sigma_x$) is applied to the qubit and trit-flip operation $\mathbb{F}_{01}$ is applied to the qutrit part of the state (\ref{eq16}) at $p=p_n$ as in Eq.~(\ref{eq9}). The Fig.~\ref{fig7} shows the non-linear curvature in  $p'$ vs.  $p~or~p_n$ in ESD (red curve) and its manipulation (green curve). The manipulation leads to avoidance for $0\leq p_n\leq 0.3586$, delay for $0.3586<p_n<0.4177$, and hastening of ESD for $0.4177<p_n<0.8452$ as green curve lies below the red curve in this range.

\subsection{$\sigma_x$ operation applied to qubit only}

\begin{figure} [H]
\centering
\includegraphics[width=0.45\linewidth]{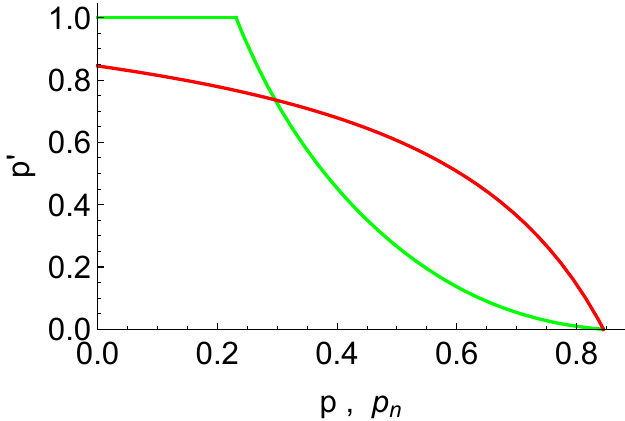}
\caption {\textit{Plot shows the non-linear curvature in  $p'$ vs.  $p ~or~p_n$ in ESD (red curve) and its manipulation (green curve) such that NOT operation is applied only on the qubit at $p=p_n$ for $x=0.5$. The action of LUOs give rise to avoidance for $0\leq p_n\leq 0.2309$, delay for $0.2309<p_n<0.2964$, and hastening of ESD for $0.2964<p_n<0.8452$ as green curve lies below the red curve in this range.}}
  \label{fig8}
\end{figure}

The NOT operation ($\sigma_x$) is applied to the qubit part of the state (\ref{eq16}) at $p=p_n$ as in Eq.~(\ref{eq9}). The Fig.~\ref{fig8} shows the non-linear curvature in  $p'$ vs.  $p~or~p_n$ in ESD (red curve) and its manipulation (green curve). The manipulation leads to avoidance for $0\leq p_n\leq 0.2309$, delay for $0.2309<p_n<0.2964$, and hastening of ESD for $0.2964<p_n<0.8452$ as green curve lies below the red curve in this range.

\subsection{$F_{01}$ applied to qutrit only}

\begin{figure} [H]
\centering
\includegraphics[width=0.45\linewidth]{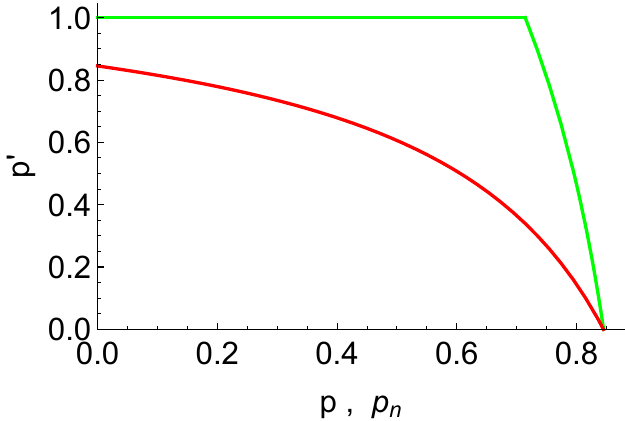}
\caption {\textit{Plot shows the non-linear curvature in  $p'$ vs.  $p ~or~p_n$ in ESD (red curve) and its manipulation (green curve) such that only trit-flip operation $\mathbb{F}_{01}$ is applied on the qutrit at $p=p_n$ for $x=0.5$. The action of LUOs give rise to avoidance for $0\leq p_n\leq 0.7143$, and delay of ESD for $0.7143<p_n<0.8452$ as green curve lies above the red curve but less than one in this range. The hastening of ESD does not occur in this case.}}
 \label{fig9}
\end{figure}

The trit-flip operation ($\mathbb{F}_{01}$) is applied only to the qutrit part of the state (\ref{eq16}) at $p=p_n$ as in Eq.~(\ref{eq9}). The Fig.~\ref{fig9} shows the non-linear curvature in  $p'$ vs.  $p~or~p_n$ in ESD (red curve) and its manipulation (green curve). The manipulation of ESD leads to avoidance for $0\leq p_n\leq 0.7143$, and delay of ESD for $0.7143<p_n<0.8452$ as green curve lies above the red curve but less than one in this range. The hastening of ESD does not occur in this case.

\subsection{$F_{02}$ applied to qutrit only}

\begin{figure} [H]
\centering
\includegraphics[width=0.45\linewidth]{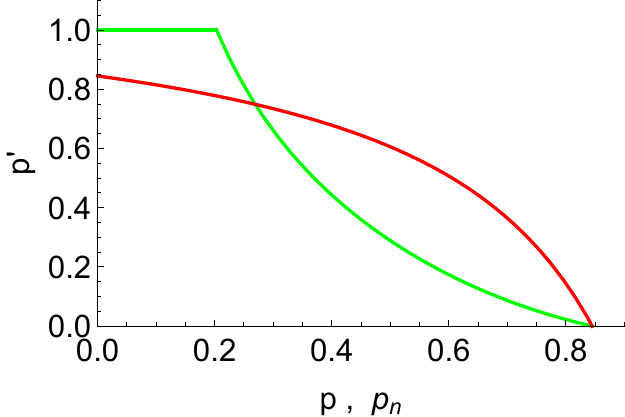}
\caption {\textit{Plot shows the non-linear curvature in  $p'$ vs.  $p ~or~p_n$ in ESD (red curve) and its manipulation (green curve) such that only trit-flip operation $\mathbb{F}_{02}$ is applied on the qutrit at $p=p_n$ for $x=0.5$.  The action of LUOs give rise to avoidance for $0\leq p_n\leq 0.2032$, delay for $0.2032<p_n<0.2693$, and hastening of ESD for $0.2693<p_n<0.8452$ as green curve lies below red curve in this range.}}
 \label{fig10}
\end{figure}

The trit-flip operation $\mathbb{F}_{02}$ is applied to the qutrit part of the state (\ref{eq16}) at $p=p_n$ as in Eq.~(\ref{eq9}). The Fig.~\ref{fig10} shows the non-linear curvature in  $p'$ vs.  $p~or~p_n$ in ESD (red curve) and its manipulation (green curve). The manipulation leads to avoidance for $0\leq p_n\leq 0.2032$, delay for $0.2032<p_n<0.2693$, and hastening of ESD for $0.2693<p_n<0.8452$ as green curve lies below the red curve in this range.

\subsection{$F_{201}$ applied to qutrit only}

\begin{figure} [h!]
  \centering
  \includegraphics[width=0.45\linewidth]{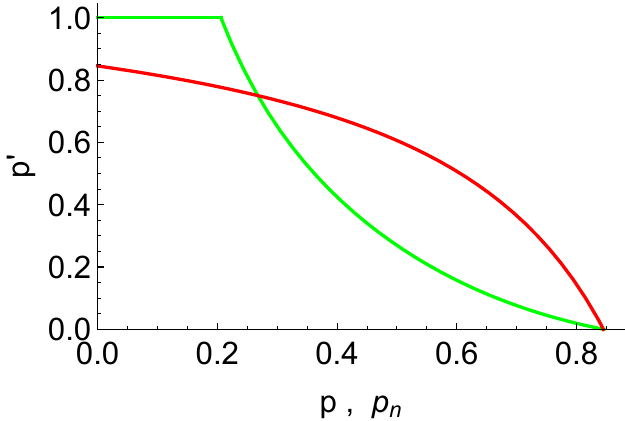}
\caption {\textit{Plot shows the non-linear curvature in  $p'$ vs.  $p ~or~p_n$ in ESD (red curve) and its manipulation (green curve) such that only trit-flip operation ($\mathbb{F}_{201}$) is applied on the qutrit at $p=p_n$ for $x=0.5$.  The action of LUOs give rise to avoidance for $0\leq p_n\leq 0.2059$, delay for $0.2059<p_n<0.2676$, and hastening of ESD for $0.2676<p_n<0.8452$ as green curve lies below the red curve in this range.}}
\label{fig11}
\end{figure}

The trit-flip operation $\mathbb{F}_{201}$ is applied to the qutrit part of the state (\ref{eq16}) at $p=p_n$ as in Eq.~(\ref{eq9}). The Fig.~\ref{fig11} shows the non-linear curvature in  $p'$ vs.  $p~or~p_n$ in ESD (red curve) and its manipulation (green curve). The manipulation leads to avoidance for $0\leq p_n\leq 0.2059$, delay for $0.2059<p_n<0.2676$, and hastening of ESD for $0.2676<p_n<0.8452$ as green curve lies below the red curve in this range.

The LUOs $\sigma_x\otimes\mathbb{F}_{102}$ and $\mathbb{I}_2\otimes\mathbb{F}_{102}$ applied on the state (\ref{eq16}) lead to same effect as $\sigma_x\otimes\mathbb{F}_{01}$ and  $\mathbb{I}_2\otimes\mathbb{F}_{01}$, respectively. The other combination of LUOs such as $\sigma_x\otimes\mathbb{F}_{02}$, and~$\sigma_x\otimes\mathbb{F}_{201}$ applied on the state (\ref{eq16}) only give rise to hastening of ESD in the entire range $0< p_n <0.8452$. 

\section{Generalization to qutrit-qutrit system}\label{S5}

In the  framework of  ADC, the dissipative  interaction of  system and
environment causes the  flow of population of the  system from excited
state to  ground state.  Therefore, any LUO which  reverses this
effect will  change the disentanglement time.  Thus, generalization of
the  above proposal  to  higher-dimensions  is  reasonably straight forward.  For example, the form  of unitary operations for two-qutrits  will be same as in  Eq.~(\ref{eq7}). However, lack of  a well  defined universal entanglement measure  for mixed  entangled states  of dimension  greater than  six, makes  it difficult  to study  the disentanglement  dynamics in  these systems because  even an  initial pure  entangled state  becomes mixed during the evolution.

For  the purpose  of  our study,  we  use Negativity  as  a witness  for entanglement, as  in general, qutrit-qutrit entanglement  is not known
to be  characterized fully, and  that negativity is a  sufficient but
not  necessary  condition  for   entanglement.   Thus,  if  negativity
undergoes  asymptotic  decay  then   this  implies  that  ESD  does not
happen. However, if negativity undergoes  sudden death (NSD), this may
be suggestive of (but does not imply) ESD.  Here, we take $a=1(p_1=ap)$
and $b=0.75~(p_2=bp)$. 

Let us consider an initially entangled two-qutrit system in the presence of ADC as given below.
\begin{equation}
\eqalign{
\rho(0)=&\frac{x}{3}(|01\rangle\langle01|+|02\rangle\langle02|+|10\rangle\langle10|+|12\rangle\langle12|+|20\rangle\langle20| 
 + |21\rangle\langle21|) \cr 
& + \frac{1-2x}{3}(|00\rangle\langle00|+ |11\rangle\langle11|+|22\rangle\langle22| 
 + |22\rangle\langle00|+ |00\rangle\langle02|),}
\label{eq17}
\end{equation}
where $0\leq x<1/3$.

For two-qutrit system, Kraus operators are given by
\begin{equation}
\mathbb{M}_{ij}=\mathbb{M}_i\otimes\mathbb{M}_j~;~i,j=0,1,2.
\label{eq18}
\end{equation}

Assuming that both the qutrits suffer identical but independent ADC, evolution of the system is given by,
\begin{equation}
\rho(p)=\sum_{i,j} \mathbb{M}_{ij} \rho(0) \mathbb{M}_{ij}^\dag.
\label{eq19}
\end{equation}

Evolution of the entanglement vs. ADC parameter ($p$) for $0\leq x<1/3$ is computed. It is found that the entangled state (\ref{eq17}) undergoes  NSD in the entire range of x-parameter for $0<x<1/3$.

Let us label another set of Kraus operators ($\mathbb{M}'_{ij}$) with the parameter $p$ replaced by $p'$ ($t$ replaced by $t'$, $t'=t-t_n$); $p'=1-\exp(-\Gamma t')$, and of the form similar to (\ref{eq18}), and apply it to the state (\ref{eq17}) to get the state of the uninterrupted system evolving in the ADC. 
\begin{equation}
\rho(p',p)=\sum_{i,j}\mathbb{M}'_{ij} \rho(0)\mathbb{M}_{ij}^{'\dag}.
\label{eq20}
\end{equation}

We choose $x=0.25$ such that the initial state (\ref{eq17}) undergoes NSD at $p=0.3636$. For $p=0$, NSD occurs at $p'=0.3636$ and for arbitrary values for $p$, NSD occurs along the non-linear curve in $pp'$ plane similar to Fig.~\ref{fig4}.

For protecting entanglement, trit-flip operation ($\mathbb{F}_{01}$) is applied to only one of the qutrits at $p=p_n$ as follows,
\begin{equation}
\rho^{(1)}(p_n)=(\mathbb{F}_{01} \otimes \mathbb{I}_3)\rho(p)(\mathbb{F}_{01} \otimes \mathbb{I}_3)^\dag.
\label{eq21}
\end{equation}

Evolution of the system afterwards in ADC is given by,
\begin{equation}
\rho^{(1)}(p',p_n)=\sum_{i,j} \mathbb{M}'_{ij}\rho^{(1)}(p_n)\mathbb{M}_{ij}^{'\dag}.
\label{eq22}
\end{equation}

\begin{figure} [H]
\begin{center}
\includegraphics[width=0.45\columnwidth]{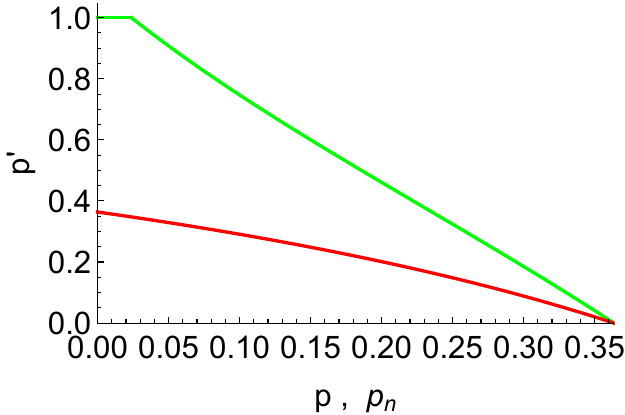}
\caption{\textit{Plot shows the non-linear curvature in  $p'$ vs.  $p ~or~p_n$ in NSD (red curve) and its manipulation (green curve) such that the trit-flip operation ($\mathbb{F}_{01}$) is applied on only one of the qutrits at $p=p_n$ for  $x=0.25$. The action of LUOs give rise to avoidance  for $0\leq p_n \leq 0.0238$, delay for $0.0238<p_n<0.3636$, and hastening of NSD does not occur in this case.}}
\label{fig12} 
\end{center}
\end{figure}

The Fig.~\ref{fig12} shows the non-linear curvature in  $p'$ vs.  $p~or~p_n$ for NSD (red curve) and its manipulation (green curve). The action of LUO ($\mathbb{F}_{01}$) gives rise to avoidance of NSD for $0\leq p_n \leq 0.0238$, delay for $0.0238<p_n<0.3636$, and hastening of NSD does not occur for this choice of parameter.

\section{Summary and discussion}\label{S6}

We have  proposed a set  of Local Unitary Operations (LUOs)  for qubit-qutrit system undergoing Entanglement  Sudden Death (ESD) in  the presence of an amplitude damping channel (ADC), such that when they are applied locally on one or both subsystems, then depending on the initial state, choice of the operation,  and its time  of application, one can  always suitably manipulate  the ESD.   We  have considered  two  different classes  of initially entangled qubit-qutrit systems which undergo ESD, and we find that  for a  given  initial  state, one  can  always  find a  suitable combination  of LUOs,  such that  when applied  at appropriate time  it can always delay  the time of ESD,  and therefore facilitate the tasks which would  not have been possible under shorter entanglement lifetime.

The  results  of   different  combinations  of  LUOs applied on  the qubit-qutrit system on  the manipulation of ESD for two different initially entangled states are summarized in the table~(\ref{tab1}) below.  In some cases, ESD can  be hastened, delayed,  as well  as avoided, whereas  in other cases,  it  can be  only  delayed  and avoided,  or  ESD  can be  only hastened.  Due to  symmetry  in the  population  of initial  entangled state,  the   NOT  operations   $\sigma_x\otimes\mathbb{F}_{102}$  and $\mathbb{I}_2\otimes\mathbb{F}_{102}$  applied  on the  either  states lead   to   same   effect  as   $\sigma_x\otimes\mathbb{F}_{01}$   and $\mathbb{I}_2\otimes\mathbb{F}_{01}$, respectively. Based on the  results in table (\ref{tab1}), the  following LUOs are advisable:  for example,  for state-II,  $\sigma_x$ on  the first qubit   and  no   action   on  the   qutrit   suffices  to   guarantee avoidance of ESD provide it is applied sufficiently early. Since this  is the simplest of all  possible combination of operations,  this  may  be  called   the  optimal  in  terms  of  gate operations.   For state-I,  all operations  allowing avoidance  are two-sided. It is  worth noting here that although the  noise is acting on both the subsystems, still  even LUO applied on only one of the  subsystems can suitably delay or avoid  the ESD as in the case of a two qubit system.

\begin{table}[h]
\begin{center}  
\renewcommand{\arraystretch}{1.25}
\begin{tabular}{ |p{3cm}||p{3cm}||p{3cm}|}
\hline
Operation  & State-I & State-II  \\
\hline \hline
 $\sigma_x \otimes \mathbb{F}_{01}$ & A, D, and H & A, D, and H   \\
  \hline
  $\sigma_x \otimes \mathbb{F}_{02}$ & only H &  only H   \\
  \hline
  $\sigma_x \otimes \mathbb{F}_{102}$ & A, D, and H &  A, D, and H   \\
  \hline
  $\sigma_x \otimes \mathbb{F}_{201}$ & only H &  Only H  \\
  \hline
  $\sigma_x \otimes \mathbb{I}_{3}$ & only H & A, D, and H  \\
  \hline
 $\mathbb{I}_{2} \otimes \mathbb{F}_{01}$ & only A and D  & only A and D  \\
  \hline
  $\mathbb{I}_{2} \otimes \mathbb{F}_{02}$ & only H &  A, D, and H   \\
  \hline
   $\mathbb{I}_{2} \otimes \mathbb{F}_{102}$ & only A and D & only A and D  \\
  \hline
   $\mathbb{I}_{2} \otimes \mathbb{F}_{201}$ & only H &   A, D, and H  \\
  \hline
\end{tabular}
\caption{\textit{Different combination of local  unitary operations applied on the  initial state-I (Eq.~\ref{eq11})  and state-II (Eq.~\ref{eq16}) resulting in avoidance (A), delay (D), and hastening (H) of ESD. Due to symmetry  in the population  of initial entangled state,  LUOs  $\sigma_x\otimes\mathbb{F}_{102}$   and
  $\mathbb{I}_2\otimes\mathbb{F}_{102}$ applied  on the  either states
  lead   to  same   effect  as   $\sigma_x\otimes\mathbb{F}_{01}$  and
  $\mathbb{I}_2\otimes\mathbb{F}_{01}$, respectively}}. \label{tab1}
\end{center}
\end{table}

Such a scheme will find application  where two parties, say Alice and
Bob, share an  entangled pair for some  quantum information processing task  and  they  know  a-priori that  ADC  is  present  in  the
environment, and therefore they can decide whether they are faced
with the prospect  of ESD.  Then, they can locally  apply suitable
LUOs at appropriate time to  delay  or  avoid  the ESD.   Our  proposed  scheme  for preserving  entanglement   longer  will   also  find   application  in entanglement distillation protocols.

 Furthermore, these LUOs can also be effective in protecting higher-dimensional entangled quantum systems in the presence of generalized amplitude damping channel as in the case of two-qubit system [\ref{24}]. Applying LUOs in the presence of generalized ADC towards the entanglement protection of higher dimensional systems could be an interesting work for future study. However, unlike ADC or generalized ADC, LUOs are not effective in protecting entanglement in the presence of bit-flip or phase-flip channel or both and the argument is as follows: under the action of ADC, population of a qubit (or qutrit) system flows from excited state to the ground state. The action of NOT operation (or more general LUO) is to reverse this effect by pumping the population from the ground to an excited state. While LUOs don’t change the amount of entanglement present in the system, they change the subsequent entanglement dynamics leading to hastening, delay or avoidance of ESD. This is the reason for effectiveness of LUOs in controlling the disentanglement dynamics in the presence of ADC. However, same reasoning doesn’t hold true for the bit-flip or phase flip channel as these are non-dissipative noises. 

Other schemes which aim to protect entanglement in a noisy environment include dynamic decoupling [\ref{56}], weak measurement (partial-collapse measurement) and quantum measurement reversal [\ref{57}], and quantum Zeno effect [\ref{58}]. Dynamic decoupling uses a sequence of $\pi$-pulses to protect the quantum states from noise. This scheme can potentially freeze the initial state and thereby preserve the quantum coherence stroboscopically to infinite time. The weak measurement and reversal scheme is based on applying a weak measurement prior to decoherence and then probabilistically reversing this operation on the decohered state. The measurement induced quantum Zeno effect [\ref{58}] for entanglement protection in cavity QED architecture, on the other hand, utilizes a very simple method of monitoring the population of the cavity modes that results in the entanglement protection exceeding its natural life-time. In our proposed method, LUOs alter the state in such a way that the decoherence effect on entanglement is minimized, i.e., for the states which were initially undergoing ESD, either time of ESD is delayed or it is averted altogether. Resource-wise, our scheme is simpler compared to the aforementioned schemes as it requires just one time intervention using LUOs but unlike the weak measurement and reversal protocol it does not restore the initial state after the decoherence.

To conclude, we have presented a scheme based on local unitary operations to protect entanglement from undergoing sudden death in the presence of amplitude damping channel for higher-dimensional systems. We have also compared and contrasted our entanglement protection scheme with other existing schemes. An interesting future line of theoretical study could be the entanglement protection in the presence of generalized amplitude damping channel in higher-dimensional systems. Our scheme will also attain more practical significance through commensurate future experiments.

\section*{Acknowledgments}\label{S7}
We thank Prof. A. R. P. Rau and Prof. R. Srikanth for useful discussions. \textit{US} would like to thank the Indian Space Research Organisation for support through the QuEST-ISRO research grant.

\section{References}\label{S8}
\begin{enumerate}
\setlength{\itemsep}{0pt}
\item \textit{W. H. Zurek 2003 Decoherence, einselection, and the quantum origins of the classical Rev. Mod. Phys. \href{https://doi.org/10.1103/RevModPhys.75.715}{ \textbf{75}, 715-775.}}\label{1}
\item \textit{R. Horodecki, P. Horodecki, M. Horodecki, K. Horodecki, 2009 Quantum entanglement Rev. Mod. Phys. \href{https://doi.org/10.1103/RevModPhys.81.865}{\textbf{ 81}, 865 (2009).}} \label{2}
\item \textit{C. H. Bennett and D. P. DiVincenzo 2000 Quantum information and computation Nature \href{https://doi.org/10.1038/35005001}{\textbf{404}, 247.}} \label{3}
\item \textit{M. A. Nielsen  and I. L. Chuang 2000 Quantum Computation and Quantum Information \href{https://doi.org/10.1017/CBO9780511976667}{Cambridge University Press, Cambridge}}. \label{4}
\item \textit{H. P. Breuer and F. Petruccione 2002 The Theory of Open Quantum Systems \href{https://doi.org/10.1093/acprof:oso/9780199213900.001.0001}{Oxford University Press, Oxford}}. \label{5}
\item \textit{T. Yu  and J. H. Eberly 2004 Finite-Time Disentanglement Via Spontaneous Emission Phys. Rev. Lett. \href{https://doi.org/10.1103/PhysRevLett.93.140404}{ \textbf{93}, 140404 (2004).}} \label{6}
\item \textit{T. Yu  and J. H. Eberly  2006 Quantum Open System Theory: Bipartite Aspects Phys. Rev. Lett. \href{https://doi.org/10.1103/PhysRevLett.97.140403}{ \textbf{97}, 140403.}} \label{7}
\item \textit{T. Yu  and J. H. Eberly 2009 Sudden Death of Entanglement Science \href{https://doi.org/10.1126/science.1167343}{\textbf{ 323} , 598.}} \label{8}
\item \textit{J. Laurat, K. S. Choi, H. Deng, C. W. Chou, H. J. Kimble 2007 Heralded Entanglement between Atomic Ensembles: Preparation, Decoherence, and Scaling Phys. Rev. Lett.  \href{https://doi.org/10.1103/PhysRevLett.99.180504}{\textbf{99}, 180504.}} \label{9}
\item \textit{M. P. Almeida, F. de Melo, M. Hor-Meyll, A. Salles, S. P. Walborn, P. H. Souto Ribeiro, L. Davidovich 2007 Environment-Induced Sudden Death of Entanglement Science \href{https://doi.org/10.1126/science.1139892}{\textbf{316}, 555.}} \label{10}
\item \textit{D. A. Lidar, I. L. Chuang, and K. B. Whaley 1998 Decoherence-free subspaces for quantum computation Phys. Rev. Lett. \href{https://doi.org/10.1103/PhysRevLett.81.2594}{\textbf{ 81}, 2594.}} \label{11}
\item \textit{Lorenza Viola, Emanuel Knill and Seth Lloyd 1999 Dynamical Decoupling of Open Quantum Systems Phys. Rev. Lett.  \href{https://doi.org/10.1103/PhysRevLett.82.2417}{\textbf{82}, 2417.}} \label{12}
\item \textit{P.W. Shor 1995 Scheme for reducing decoherence in quantum computer memory Phys. Rev. A \href{https://doi.org/10.1103/PhysRevA.52.R2493}{\textbf{52}, R2493.}} \label{13}
\item \textit{A.M. Steane 1996 Error Correcting Codes in Quantum Theory Phys. Rev. Lett. \href{https://doi.org/10.1103/PhysRevLett.77.793}{\textbf{77}, 793.}} \label{14}
\item \textit{P. Facchi, D. A. Lidar, and S. Pascazio 2004 Unification of dynamical decoupling and the quantum Zeno effect Phys. Rev. A \href{https://doi.org/10.1103/PhysRevA.69.032314}{\textbf{69}, 032314.}} \label{15}
\item\textit{Naoki Yamamoto, Hendra I. Nurdin, Matthew R. James, and Ian R. Petersen 2008 Avoiding entanglement sudden death via measurement feedback control in a quantum network Phys. Rev. A  \href{https://doi.org/10.1103/PhysRevA.78.042339}{\textbf{78}, 042339.}} \label{16}
\item \textit{ S. Maniscalco, F. Francica, R. L. Zaffino, N. L. Gullo, and F. Plastina 2008 Protecting entanglement via the quantum Zeno effect Phys. Rev. Lett. \href{https://doi.org/10.1103/PhysRevLett.100.090503}{\textbf{100}, 090503.}} \label{17}
\item \textit{J. G. Oliveira, Jr., R. Rossi, Jr., and M. C. Nemes 2008 Protecting, enhancing, and reviving entanglement Phys. Rev. A  \href{https://doi.org/10.1103/PhysRevA.78.044301}{\textbf{78}, 044301.}} \label{18}
\item \textit{Q. Sun, M. Al-Amri, L. Davidovich, and M. S. Zubairy 2010 Reversing entanglement change by a weak measurement Phys. Rev. A\href{https://doi.org/10.1103/PhysRevA.82.052323}{\textbf{ 82}, 052323.}}\label{19}
\item \textit{A. N. Korotkov and K. Keane 2010 Decoherence suppression by quantum measurement reversal Phys. Rev. A \href{https://doi.org/10.1103/PhysRevA.81.040103}{\textbf{81},040103(R).}}\label{20}
\item \textit{A.R.P. Rau, M. Ali, and G. Alber 2008 Hastening, delaying or averting sudden death of quantum entanglement EPL \href{https://doi.org/10.1209/0295-5075/82/40002}{\textbf{82}, 40002.}} \label{21}
\item \textit{Ashutosh Singh, Siva Pradyumna, A R P Rau, and Urbasi Sinha 2017 Manipulation of entanglement sudden death in an all-optical setup J. Opt. Soc. Am. B \href{https://doi.org/10.1364/JOSAB.34.000681}{\textbf{34}, 681-690.}} \label{22}
\item\textit{Mahmood Irtiza Hussain, Rabia Tahira and Manzoor Ikram 2011 Manipulating the Sudden Death of Entanglement in Two-qubit Atomic Systems J. Korean Phys. Soc. \href{https://doi.org/10.3938/jkps.59.2901}{\textbf{59}, 4, pp. 2901-2904.}} \label{23}
\item \textit{M. Ali, A. R. P. Rau, and G. Alber 2009 Manipulating entanglement sudden death of two-qubit X-states in zero- and finite-temperature reservoirs J. Phys. B: At. Mol. Opt. Phys. \href{https://doi.org/10.1088/0953-4075/42/2/025501} {\textbf{ 42}, 025501(8).}}\label{24}
\item \textit{P. G. Kwiat, A. J. Berglund, J. B. Altepeter, and A. G. White 2000 Experimental verification of decoherence-free subspaces Science \href{https://doi.org/10.1126/science.290.5491.498} {\textbf{ 290}, 498.}} \label{25}
\item \textit{D. Kielpinski, V. Meyer, M. A. Rowe, C. A. Sackett, W. M. Itano, C. Monroe, and D. J. Wineland 2001 A decoherence-free quantum memory using trapped ions Science \href{https://doi.org/10.1126/science.1057357}{\textbf{ 291}, 1013.}}\label{26}
\item \textit{L. Viola, E. M. Fortunato, M. A. Pravia, E. Knill, R. Laflamme, and D. G. Cory 2001 Experimental realization of noiseless subsystems for quantum information processing Science \href{https://doi.org/10.1126/science.1064460}{\textbf{293}, 2059.}}\label{27}
\item \textit{Michael J. Biercuk, Hermann Uys, Aaron P. VanDevender, Nobuyasu Shiga, Wayne M. Itano \& John J. Bollinger 2009 Optimized dynamical decoupling in a model quantum memory Nature  \href{https://doi.org/10.1038/nature07951}{\textbf{458}, 996.}} \label{28}
\item\textit{Jiangfeng Du, Xing Rong, Nan Zhao, Ya Wang, Jiahui Yang \& R. B. Liu 2009 Preserving electron spin coherence in solids by optimal dynamical decoupling Nature \href{https://doi.org/10.1038/nature08470}{ \textbf{461}, 1265.}} \label{29}
\item \textit{ J.C. Lee, Y.C. Jeong, Y.S. Kim, and Y.H. Kim 2011 Experimental demonstration of decoherence suppression via quantum measurement reversal Opt. Express  \href{https://doi.org/10.1364/OE.19.016309}{\textbf{19}, 16309 - 16316.}} \label{30}
\item \textit{Y.S. Kim, J.C. Lee, O. Kwon, and Y.-H. Kim 2012 Protecting entanglement from decoherence using weak measurement and quantum measurement reversal Nature Phys. \href{https://doi.org/10.1038/nphys2178}{\textbf{8}, 117.}} \label{31}
\item \textit{H.T. Lim, J.C. Lee, K.H. Hong, and Y.H. Kim 2014 Avoiding entanglement sudden death using single-qubit quantum measurement reversal Opt. Express \href{https://doi.org/10.1364/OE.22.019055}{\textbf{22}, 19055 (2014).}} \label{32}
\item \textit{Jong-Chan Lee, Hyang-Tag Lim, Kang-Hee Hong, Youn-Chang Jeong, M.S. Kim \& Yoon-Ho Kim 2014 Experimental demonstration of delayed-choice decoherence suppression Nature Communications \href{https://doi.org/10.1038/ncomms5522}{\textbf{5}, 4522.}} \label{33}
\item \textit{Jin-Shi Xu, Chuan-Feng Li, Ming Gong, Xu-Bo Zou, Cheng-Hao Shi, Geng Chen, and Guang-Can Guo 2010 Experimental demonstration of photonic entanglement collapse and revival Phys. Rev. Lett. \href{https://doi.org/10.1103/PhysRevLett.104.100502}{\textbf{ 104}, 100502.}} \label{34}
\item \textit{Erhard M., Krenn M. \& Zeilinger A. 2020 Advances in high-dimensional quantum entanglement Nat Rev Phys \href{https://doi.org/10.1038/s42254-020-0193-5}{\textbf{2}, 365–381.}}\label{35}
\item \textit{Daniel Collins, Nicolas Gisin, Noah Linden, Serge Massar, and Sandu Popescu 2002 Bell inequalities for arbitrarily high-dimensional systems Phys. Rev. Lett. \href{https://doi.org/10.1103/PhysRevLett.88.040404}{\textbf{88}, 040404.}}\label{36}
\item \textit{Ann Kevin, and Jaeger Gregg 2008 Entanglement sudden death in qubit-qutrit systems Phys. Lett. A \href{https://doi.org/10.1016/j.physleta.2007.07.070} {\textbf{372}, 579-583.}} \label{37}
\item \textit{ M. Ali, A R P Rau, and Kedar Ranade 2007 Disentanglement in qubit-qutrit systems arXiv \href{https://arxiv.org/abs/0710.2238}{0710.2238.}} \label{38}
\item \textit{Mazhar Ali 2009 Quantum Control of Finite-time Disentanglement in Qubit-Qubit and Qubit-Qutrit Systems \href{http://tuprints.ulb.tu-darmstadt.de/id/eprint/1895}{Ph.D. Thesis.}}\label{39}
\item \textit{Xing Xiao 2008 Protecting qubit-qutrit entanglement from amplitude damping decoherence via weak measurement and reversal Phys. Scr. \href{http://dx.doi.org/10.1088/0031-8949/89/6/065102}{\textbf{89}, 065102.}}\label{40}
\item\textit{X. Xiao and Y. L. Li 2013 Protecting qutrit-qutrit entanglement by weak measurement and reversal Eur. Phys. J. D \href{http://dx.doi.org/10.1140/epjd/e2013-40036-3}{\textbf{67}, 204.}}\label{41} 
\item\textit{Xiang-Ping Liao and Mao-Fa Fang and Man-Sheng Rong and Xin Zho 2017 Protecting free-entangled and bound-entangled states in a two-qutrit system under decoherence using weak measurements Journal of Modern Optics \href{https://doi.org/10.1080/09500340.2016.1271149}  {\textbf{64}, 12, pp. 1184-1191.}}\label{42}
\item \textit{Mazhar Ali 2010 Distillability sudden death in qutrit-qutrit systems under amplitude damping J. Phys. B  \href{https:/doi.org/10.1088/0953-4075/43/4/045504}{\textbf{43}, 045504.}}\label{43}
\item \textit{L. Derkacz and L. Jakobczyk 2006 Quantum interference and evolution of entanglement in a system of three-level atoms Phys. Rev. A \href{https://doi.org/10.1103/PhysRevA.74.032313} {\textbf{74}, 032313.}}\label{44}
\item \textit{A. Peres 1996 Separability Criterion for Density Matrices Phys. Rev. Lett. \href{https://doi.org/10.1103/PhysRevLett.77.1413}{\textbf{77}, 1413.}}\label{45} 
\item \textit{M.Horodecki, P. Horodecki  and R. Horodecki 1996 Separability of mixed states: necessary and sufficient conditions Phys. Lett. A \href{https://doi.org/10.1016/S0375-9601(96)00706-2}{\textbf{223}, 1.}} \label{46}
\item \textit{Paweł Horodecki, Maciej Lewenstein, Guifré Vidal, and Ignacio Cirac 2000 Operational criterion and constructive checks for the separability of low-rank density matrices Phys. Rev. A \href{https://link.aps.org/doi/10.1103/PhysRevA.62.032310}{\textbf{62}, 032310.}} \label{47}
\item \textit{O. Rudolph  2004 Computable Cross-norm Criterion for Separability Letters in Mathematical Physics \href{https://doi.org/10.1007/s11005-004-0767-7}{\textbf{70}, 57–64}}.\label{48}
\item \textit{O. Rudolph 2005 Further results on the cross norm criterion for separability Quant. Inf. Proc. \href{https://10.1007/s11128-005-5664-1}{\textbf{4},  3, pp. 219 - 239}}.\label{49}
\item \textit{Kai Chen, Ling-An Wu 2003 A matrix realignment method for recognizing entanglement Quantum Inf. Comput. \href{http://dl.acm.org/citation.cfm?id=2011534.2011535}{\textbf{3}, 3, pp.193-202.}} \label{50}
\item \textit{E. Hagley, X. Ma$\hat{\iota}$tre, G. Nogues, C. Wunderlich, M. Brune, J. M. Raimond, and S. Haroche 1997 Generation of Einstein-Podolsky-Rosen Pairs of Atoms Phys. Rev. Lett. \href{https://doi.org/10.1103/PhysRevLett.79.1}{\textbf{79}, 1.}}  \label{51}
\item \textit{B. P. Lanyon, T. J. Weinhold, N. K. Langford, J. L. O’Brien, K. J. Resch, A. Gilchrist, and A. G. White 2008 Manipulating Biphotonic Qutrits Phys. Rev. Lett. \href{https://doi.org/10.1103/PhysRevLett.100.060504}{\textbf{100}, 060504.}}\label{52}
\item \textit{Robert Fickler, Radek Lapkiewicz, William N. Plick, Mario Krenn, Christoph Schaeff, Sven Ramelow, Anton Zeilinger 2012 Quantum Entanglement of High Angular Momenta Science \href{https://doi.org/10.1126/science.1227193}{\textbf{338}, 640-643.}} \label{53}
\item \textit{E. J. Galvez, S. M. Nomoto, W. H. Schubert, and M. D. Novenstern 2011 Polarization-Spatial-Mode Entanglement of Photon Pairs International Conference on Quantum Information, OSA Technical Digest (CD) (Optical Society of America), paper QMI18.} \label{54}
\item \textit{Zheng Shi-Biao 2006 Production of Entanglement of Multiple Three-Level Atoms with a Two-Mode Cavity Commun. Theor. Phys. \href{https://doi.org/10.1088/0253-6102/45/3/031} {\textbf{45}, 539-541.}} \label{55}
\item\textit{ L. Viola, E. Knill, and S. Lloyd 1999 Dynamical Decoupling of Open Quantum Systems Phys. Rev. Lett.  \href{https://doi.org/10.1103/PhysRevLett.82.2417}{\textbf{82}, 2417.}}\label{56}
\item \textit{Yong-Su Kim, Jong-Chan Lee, Osung Kwon and Yoon-Ho Kim 2012 Protecting entanglement from decoherence using weak measurement and quantum measurement reversal Nature Physics \href{https://doi.org/10.1038/NPHYS2178}{\textbf{8}, 117–120.}} \label{57}
\item \textit{S. Maniscalco, F. Francica, R. L. Zaffino, N. L. Gullo, and F. Plastina 2008 Protecting entanglement via the quantum Zeno effect, Phys. Rev. Lett. \href{https://doi.org/10.1103/PhysRevLett.100.090503}{\textbf{100}, 090503}}. \label{58}
\end{enumerate}
\end{document}